\documentclass[a4paper,12pt]{article}

\usepackage{amsmath}
\usepackage{amssymb}
\usepackage{latexsym}
\usepackage{cite}
\usepackage[dvips]{graphicx}
\usepackage{bbold}

\newcommand{\be}{\begin{equation}}
\newcommand{\ee}{\end{equation}}
\newcommand{\half}{\frac{1}{2}}

\newcommand{\ka}{\kappa}

\newcommand{\mcA}{{\mathcal A}}

\newcommand{\mcG}{{\mathcal G}}
\newcommand{\mcL}{{\mathcal L}}
\newcommand{\mcN}{{\mathcal N}}

\newcommand{\mcV}{{\mathcal V}}
\newcommand{\mcW}{{\mathcal W}}

\def\beq{\begin{equation}}
\def\eeq{\end{equation}}

\def\ka{\kappa}

\def\hs{\hspace}
\def\vs{\vspace}

\begin{document}

\begin{titlepage}
\begin{flushright}
HD-THEP-04-42\\
CERN-PH-TH/2004-211\\
October 28, 2004
\end{flushright}
\vspace{0.6cm}
\begin{center}
{\Large \bf (BPS) Fayet-Iliopoulos Terms in 5D Orbifold SUGRA} 
\end{center}
\vspace{0.5cm}

\begin{center}
{\large   
Filipe Paccetti Correia$^{a1}$,
Michael G. Schmidt$^{a2}$,
Zurab Tavartkiladze$^{b3}$}

\vspace{0.3cm}

{\em 
${}^a$ Institut f\"ur Theoretische Physik,
Universit\"at Heidelberg\\ 
Philosophenweg 16, 69120 Heidelberg, Germany}\\
{\em 
${}^b$ Physics Department, Theory Division, CERN, CH-1211 Geneva 23, Switzerland}
\end{center}
\vspace{0.4cm}
\begin{abstract}
We discuss Fayet-Iliopoulos terms in the context of five-dimensional supergravity compactified on an orbifold. For this purpose we use our superfield formulation of the off-shell 5D SUGRA. In the case of tuned FI terms, contrary to other claims, we find BPS solutions which ensure that N=1 supersymmetry is unbroken also in warped geometries. As in the rigid case, the FI terms induce odd masses for charged hypermultiplets, leading to the (de)localisation of the KK wave-functions near the fix-point branes. In the case of ungauged U(1)$_R$ symmetry, we present also supersymmetric warped solutions in the presence of non-trivial profiles of charged hyperscalars. 
\end{abstract} 
\vspace{8cm}
\footnoterule

{\small
\noindent$^1 $E-mail address: f.paccetti@thphys.uni-heidelberg.de\\  \noindent$^2 $E-mail address: m.g.schmidt@thphys.uni-heidelberg.de\\
\noindent$^3 $E-mail address: Zurab.Tavartkiladze@cern.ch} 

%
%

\end{titlepage}

\section{Introduction}\label{sec:intro}
In this letter we present a discussion of Fayet-Iliopolous terms within 5D supergravity compactified on the $S^1/{\mathbb Z}_2$ orbifold. When FI terms where considered first, in the context of 4D supersymmetric theories \cite{fayet}, they were seen as a means of breaking supersymmetry and/or gauge symmetry. Later, their utmost relevance for cosmology was also recognized, as it became clear that they could be at the origin of de Sitter configurations, and more generally of inflationary scenarios \cite{binetruy96}. While in global (4D) supersymmetric theories the introduction of FI terms is rather straightforward, it turns out that in supergravity this is not the case. In fact, the compatibility of local supersymmetry and FI terms requires the U(1) gauge symmetry in question to be an $R$-symmetry \cite{freedman76,barbieri82,binetruy04}, and therefore the gravitino has to be charged. In addition, they only can be radiatively generated in the presence of a mixed U(1)-gravitational anomaly.  

In five-dimensional orbifolds the situation gets another twist. In the rigid case, the FI terms can be consistently introduced at the 4D fix-point branes, but unlike in the 4D case they can be tuned in such a way that neither supersymmetry nor the U(1) gauge symmetry are broken \cite{ghilen01,barbie02,groot02,martiFI}. As it was pointed out in \cite{barbie02}, FI terms can be generated radiatively even in the case that the mixed anomaly is absent, but turn out to be of the \emph{tuned} type that we just mentioned. The effect of such tuned FI terms is to induce a stepwise VEV of the (odd) scalar component of the U(1) vector multiplet, which leads to the localisation of zero-modes of charged hypermatter \cite{groot02,martiFI}. On the other hand, if this U(1) symmetry is part of a larger bulk gauge symmetry $\mcG$, the VEV of the vector scalar will break $\mcG$ in the bulk while orbifolding breaks it at the boundary. The relevance of this for \emph{calculable} power-law unification has been recently emphasized in \cite{hebecker04}.         

A discussion of the embedding of (tuned) FI terms in 5D orbifold supergravity was first given in ref.\cite{barbie02}, where it was pointed out that they are associated with a bulk Chern-Simons term with one U(1)$_{FI}$ gauge boson and two graviphotons. In particular, the strength of the FI terms is fixed by the strength of the stepwise coupling of the CS term. As in the rigid case, the tuned FI terms lead to a stepwise VEV of the vector scalar, and therefore to the localisation of charged hypermultiplets. This analysis was recently extended in \cite{abe04} to orbifold SUGRA with warped geometry, i.e. these authors considered the possibility of gauging the U(1)$_R$ symmetry. They came to the, in our view, incorrect conclusion that in the presence of a warped geometry, unless hypermatter is introduced, SUSY is broken by non-vanishing (tuned) FI terms.

We will here show that tuned FI terms don't lead to the breaking of $\mcN=1$ supersymmetry, even in a warped geometry. In other words, we will see that the BPS conditions have solutions in the presence of tuned FI terms, even if we gauge the U(1)$_R$ symmetry. For this reason we call this type of FI terms \emph{BPS} FI terms. To obtain the BPS conditions, we use the superfield approach to 5D supergravity, that we recently have presented in \cite{us04} (see also the subsequent work \cite{abesaka04}), based on the work of Fujita, Kugo and Ohashi on off-shell 5D conformal SUGRA in component form \cite{kugo,kugo02}. As we have already shown in \cite{us04}, in this formalism both the gauging of the U(1)$_R$ and the introduction of the BPS FI terms, which are obtained by the introduction of \emph{stepwise} couplings, can be consistently made without having to rely on the 4-form mechanism of ref.\cite{ber}. In fact, the stepwise couplings introduced directly in the superspace action give rise to the correct brane-localized couplings upon suitable partial integrations. In addition, the BPS conditions correspond to the conditions of D-flatness and F-flatness, which as we will see are rather simple to write down within the superfield formalism. 
 
We consider here two different cases, namely with and without charged hypermultiplets, and in both cases we find SUSY vacua. While in absence of hypermultiplets we obtain a solution with a warp-factor of the Randall-Sundrum type\cite{randall99} and a stepwise VEV for the vector scalar, the inclusion of two bulk hypermultiplets with opposite U(1)$_{FI}$ charges allows for more general solutions. In particular, in the case the U(1)$_R$ is not gauged, we obtain warped solutions corresponding to the presence of negative brane tensions. These are induced by non-vanishing profiles of the two even hyperscalars, which are localised near opposite branes.

\section{BPS FI Terms}\label{sec:BPS_FI_Terms}

Before we discuss the 5D orbifold SUGRA case let us shortly review the status of FI terms in the rigid case. In 4D they are allowed (for Abelian gauge groups) and cause either the breaking of SUSY or of the corresponding U(1)$_{FI}$ gauge symmetry. In 5D orbifolds, the situation is different \cite{ghilen01,barbie02,groot02,martiFI} due to the existence of the 4D chiral superfield $\Sigma=\tfrac{1}{2}(M+iA_y)+\cdots$ (we take $e^5_y=1$), which accompanies the 4D vector superfield $V$. Indeed, the derivative $\partial_y\Sigma$ can cancel the FI terms localized at the fixed point boundaries, in which case SUSY remains unbroken and $M$ gets a stepwise VEV. This cancelation takes only place in case the FI terms in the two boundaries are \emph{tuned}, having opposite signs and equal absolute values at different branes. Using the superfield description of 5D rigid supersymmetry \cite{arkani01,marti01,hebecker01,dudas04}, these FI terms can be writen as
\be
          \mcL_{FI}=-4[\delta(y)-\delta(y-\pi R)]\int d^4\theta \xi V.       
\ee
We now make the observation that in the rigid case the (tuned) FI term can be rewriten as follows
\be\label{eq:FIrigid}
         \mcL_{FI}=-2\int d^4\theta \xi (\partial_y\epsilon (y)) V=2\int d^4\theta\xi\epsilon (y)[\partial_y V-(\Sigma+\Sigma^+)]=-2\int d^4\theta \,\xi\epsilon (y)\mcV_y,
\ee
where we introduced the gauge invariant $\mcV_y\equiv\Sigma+\Sigma^+-\partial_y V$. There is also a term in the Lagrangian, quadratic in $\mcV_y$, which is responsible for part of the kinetic terms \cite{arkani01}:
\be
                        \mcL\supset\int d^4\theta \,(\mcV_y)^2.
\ee
This can be combined with eq.\eqref{eq:FIrigid} to get
\be
                  \mcL\supset\int d^4\theta \,(\mcV_y-\xi\epsilon (y))^2.
\ee
From this expression it becomes clear that the only effect of the FI terms is to shift the lowest component of $\Sigma$ as $M\to M+\xi\epsilon(y)$, which doesn't break SUSY. The U(1)$_{FI}$ is also unbroken since $\Sigma$ is neutral under this group.

\paragraph{5D orbifold SUGRA and FI terms.} In our study we will use our superfield approach to 5D orbifold SUGRA \cite{us04}. For the sake of brevity, we will here only recall the results we need and refer the reader to that work for more details. We assume in the following that the metric is of the warped type, i.e.
\be
                        ds^2=e^{2\sigma(y)}\eta_{\mu\nu}dx^{\mu}dx^{\nu}-(e^5_y)^2 dy^2,
\ee
where the f\"unfbein's component $ e^5_y$ can also be $y$-dependent. Eventually, we will later on choose the gauges $e^5_y=e^{-2\sigma}$ or $e^5_y=\textup{const.}$ for practical reasons. Note that the warp factor $\sigma (y)$ is not fixed \emph{a priori} but will be determined from the equations of motion.

The off-shell description of 5D supergravity coupled to $n_V$ physical Abelian vector multiplets requires the introduction of $n_V+1$ off-shell vector mutiplets, ${\mathbb V}^I$ ($I=0,\dots,n_V$), connected by constraints to be described below \cite{kugo}. One of the $n_V+1$ gauge bosons will become the graviphoton. Each of the 5D off-shell vector multiplets corresponds to a vector superfield $V^I$ and a chiral superfield $\Sigma^I$ of $\mcN=1$ SUSY:
\be 
                   V^I=-\theta\sigma^{\mu}{\bar\theta}\,e^{\sigma}A_{\mu}^I+\theta^2{\bar\theta}\,e^{3\sigma/2}2i{\bar\omega}^{2I}-{\bar\theta}^2\theta\,e^{3\sigma/2}2i\omega^{2I}+\half\theta^2{\bar\theta}^2\, e^{2\sigma}D^I,
\ee
\be
                   \Sigma^I=\half(e^5_yM^I+iA_y^I)+\theta\, e^{\sigma/2}2(i e^5_y\omega^{1I}+\ka M^I\psi_y^1)+\theta^2 e^{\sigma}F^I_{\Sigma}.
\ee
Here $M^I$ is the scalar component of the vector multiplet, $A_{\mu}^I$ and $A_y^I$ are the 4D and fifth components of the gauge boson, and the 2-component spinors $(\omega^{1I},\omega^{2I})$ arise from the 5D gaugino, in the same way as $(\psi_y^1,\psi_y^2)$ from the 5th component of the gravitino. (The auxiliary fields $F^I_{\Sigma}$ and $D^I$ can also be written as combinations of fields of the 5D off-shell SUGRA of Fujita, Kugo and Ohashi \cite{kugo02,us04}.) As we stated above, the components of the 5D vector multiplets are connected by two constraints, namely
\be
                  \mcN(M)=\ka^{-2},\qquad \mcN_I(M)\omega^I=0,
\ee 
where the \emph{norm function} $\mcN(M)$ is a cubic function of the vector scalars:
\be
                   \mcN(M)=\ka c_{IJK}M^I M^J M^K,          
\ee  
and the $c_{IJK}$ are symmetric real coefficients. (The gravity coupling $\ka$ is related to the 5D Planck mass by $\ka=(M_5)^{-3/2}$.)

We must consider also another (even) superfield, ${\mathbb W}_y$, which contains elements associated with the so-called radion superfield. It is given by:
\be
                 {\mathbb W}_y=e^{-\sigma}e^5_y+\theta\,e^{-\sigma/2}2\ka\psi^1_y+{\bar\theta}\,e^{-\sigma/2}2\ka{\bar\psi}^1_y+\cdots
\ee
The gauge invariant superfield $\mcV_5$, that we introduced above for the rigid case, becomes now
\be
                 \mcV_5\equiv \frac{\Sigma+\Sigma^+-\partial_y V}{{\mathbb W}_y}+\cdots,
\ee
where the dots stay for terms involving odd components of the 5D Weyl multiplet which are here set to zero. Note that the ${\mathbb W}_y$ term in the denominator, which involves the 5th component of the gravitino, is necessary to ensure invariance of $\mcV_5$ under local supersymmetry.

In terms of the superfields we have just introduced, the vector part of the Lagrangian reads \cite{us04},
\be\begin{split}\label{eq:lagrange_vec}
                 \mcL_V=&\frac{1}{4}\int d^2\theta \,\left(-\mcN_{IJ}(\Sigma)\mcW^{\alpha I}\mcW_{\alpha}^J+\frac{1}{12}\mcN_{IJK}{\bar D}^2(V^ID^{\alpha}\partial_yV^J-D^{\alpha}V^I\partial_yV^J)\mcW^K_{\alpha}\right)\\
                        &\qquad+\textup{h.c.} -\int d^4\theta \,{\mathbb W}_y\,\mcN ({\mcV}_5).
\end{split}\ee
Note that here the norm function $\mcN$, which was earlier defined as a cubic function of the vector scalars $M^I$, plays now the r\^ ole of a prepotential, and is to be seen as a function of its argument (for instance $\mcN(\mcV_5)=\ka c_{IJK}\mcV_5^I\mcV_5^J\mcV_5^K$).

\vspace{12pt}
We now argue that in the case of SUGRA, a term similar to expression \eqref{eq:FIrigid} is obtained with a \emph{norm}-function of the following form (proposed in \cite{barbie02})
\be\label{eq:norm}
         \ka^{-1}\mcN (M)=(M^0)^3-M^0 (M^1)^2+2\ka\xi\epsilon (y) (M^0)^2 M^1,
\ee
where, to ensure that $\mcN$ has even orbifold parity, $M^0$ and $M^1$ must have positive and negative parities, respectively. It is not hard to see that the last term in the norm-function contributes to the Lagrangian a term  (see eq.\eqref{eq:lagrange_vec})
\be\label{eq:vectorFI}
      -\int d^4\theta \,{\mathbb W}_y\,\mcN ({\mcV}_5)\supset -2{(\ka M^0)^2}\int d^4\theta\, e^5_y e^{\sigma}\,\xi\epsilon (y)  {\mcV}_5^1,
\ee
which indeed has the same form as the tuned FI term in rigid SUSY but also takes into account the warped geometry. One sees that here it is the vector multiplet ${\mathbb V}^{I=1}$ which gauges the U(1)$_{FI}$ symmetry, for which there are FI terms. Due to its orbifold parity, brane localized FI terms involving $V^0$ are not possible.

In addition to the 5D multiplets that we presented already, one can introduce also bulk hypermultiplets, both physical and compensator ones. Note that at least one compensator hypermultiplet is required, to get a sensible theory. In this section we will consider the case with no physical hypermultiplets, and only one compensator multiplet. The compensator hypermultiplet corresponds to a pair of chiral superfields $(h,~h^c)$, where we take $h$ to have positive orbifold parity, ${h}^c$ to have negative. We have
\be
                   h=e^{3\sigma/2}\ka^{-1}+\theta^2\,e^{5\sigma/2}F_h,\qquad h^c=\theta^2\,e^{5\sigma/2}F_h^c.
\ee

We will gauge an U(1)$_R$ subgroup of the SU(2)$_R$ by coupling the compensator hypermultiplet $(h,~h^c)$ to the ${\mathbb V}^0$ vector multiplet with an \emph{odd} gauge coupling, $g_0\epsilon(y)$, as in \cite{us04}. The D-term Lagrangian does not only arises from eq.\eqref{eq:lagrange_vec}, but also has a contribution from the compensator Lagrangian
\be
                    \mcL_{comp}= -2\int d^4\theta \,{{\mathbb W}}_y\,({ h}^+ e^{-g_0\epsilon (y) {V^0}}{ h}+h^{c+}e^{g_0\epsilon (y) V^0} h^c)-2\left(\int d^2\theta \,h^c(\partial_y -g_0\epsilon(y)\Sigma^0)h+\textup{h.c.}\right).
\ee
The total D-term Lagrangian is thus
\be
                \mcL_D=e^{4\sigma}e^5_y\left [ -\frac{1}{4}\mcN_{IJ}(M) D^I D^J-\frac{e^{-2\sigma}e^y_5}{2} (\partial_y e^{2\sigma}\mcN_I(M))D^I+ M_5^3 g_0 \epsilon (y) D^0  \right].
\ee

As it was pointed out in \cite{us04}, the BPS conditions are that the F-terms and D-terms vanish. In particular we must have $D^I=0$. Now, it follows from the Lagrangian above that 
\be
                D^I=\mcN^{IJ}\left(2M_5^3 g_0 \epsilon (y) \delta_J^0 - e^{-2\sigma} e^y_5 \partial_y e^{2\sigma}\mcN_J\right),
\ee
and so the BPS condition $D^I=0$ becomes 
\be\label{eq:BPS_D_without}
      \partial_y (e^{2\sigma}\mcN_J)=  2 e^{2\sigma}e^5_y\,M_5^3  \epsilon (y) g_0\delta_J^0.       
\ee 
Since $\mcN_1$ has negative parity, the BPS equation with $I=1$ is solved by
\be
                   \mcN_1=0\Rightarrow -2M^0\, M^1+2{\ka}\xi\epsilon(y)(M^0)^2=0,
\ee
that is \cite{us04}
\be\label{eq:VevM^1}
                    M^1=\ka\xi\epsilon(y)M^0.
\ee
The value of $M^0$ then follows readily from $\mcN=\ka^{-2}$, being
\be\label{eq:VevM^0}
                    M^0=M_5^{3/2}(1+(\ka\xi)^2)^{-1/3}.
\ee
Finally, the metric is obtained by solving the BPS equation with $I=0$. In the gauge $e^5_y=e^{-2\sigma}$, we obtain
\be
                    e^{2\sigma}\mcN_0=t_0+2g_0 M_5^3 |y|,
\ee
where $t_0$ is an integration constant. We get
\be
                    e^{2\sigma}=\frac{M^0}{3M_5^{3}}[t_0+2g_0 M_5^3 |y|].
\ee
If prefered, one can introduce a new coordinate $z$ defined by $dz=e^{-2\sigma(y)}dy$. In terms of this variable the metric becomes
\be
                 ds^2=e^{2\sigma}dx^2-dz^2, \quad\textup{with}\quad e^{2\sigma(z)}=\exp{\left(\tfrac{2}{3}g_0M^0|z|\right)}.
\ee
One notices that since $M^0$ decreases with increasing $\xi$, a non-vanishing FI term has the effect of reducing the warping of the geometry.

It is clear from this discussion that the presence of the FI terms, even in a warped geometry, doesn't lead to supersymmetry breaking, due to the fact that the \emph{odd} scalar $M^1$ \emph{absorbs} the FI term, just as in the case of rigid SUSY.\footnote{The authors of ref.\cite{abe04} obtain the opposite result. The point is that these authors introduce an odd scalar field $\phi$ to parametrise the very special manifold defined by $\mcN(M)=\ka^{-2}$. The $M^J$ are then functions of $\phi$, but the relation between $M^1$ and $\phi$ also involves $\epsilon(y)$. This means that
\be
                      \partial_y M^1= \frac{\partial M^1}{\partial \phi}\partial_y \phi+\frac{\partial M^1}{\partial \epsilon}\partial_y\epsilon (y),\nonumber
\ee
but in \cite{abe04} the second term on the r.h.s. was neglected, e.g. in going from the third equation in eqs.(22) to the third equation in eqs.(23) of \cite{abe04}.} 

One can consider the warped geometry without the SUGRA setting just by taking $M^0$ and the warp-factor as constant backgrounds. Then the solution for $M^1$ still would be dictated from the condition of unbroken SUSY. However, without the SUGRA (which is gauged) the relations between the bulk cosmological constant and the brane tensions are assumed ad hoc. Also, in the rigid limit there is no BPS equation which gives the solution for the warp-factor.  

Before we close this section let us point out that the above results are robust against radiative corrections. In fact, the form of the tree-level FI term obtained from the norm function \eqref{eq:norm}, $\xi(z)\equiv \ka\xi\,e^{-2\sigma}\partial_z \{e^{2\sigma}\epsilon(z)(M^0)^2\}$, is compatible with the 1-loop result obtained in the rigid SUSY case\footnote{We use here the 3rd version of \cite{hira}, in particular its eq.(3.12). This version of \cite{hira} differs from previous ones notably in the use of a position-dependent cut-off. Their $\xi^{1loop}(z)$ is obtained from ours by multiplying \eqref{eq:hirayo} with a factor of $e^{2\sigma}$ and using $m_i=c_i k$. The last term in eq.\eqref{eq:hirayo} differs by a factor of 2. To obtain eq.\eqref{eq:hirayo} we summed over a non-anomalous bulk field content.}
\be\label{eq:hirayo}
                      \xi^{1loop}(z)=\frac{\Lambda}{16\pi^2}\sum_i q_i m_i\left [ (\delta(z)-\delta(z-\pi R))+k\right],
\ee
where $k\equiv\partial_z\sigma$. Indeed, in the rigid limit we have $\xi(z)\to 2\ka\xi(M^0)^2\{\delta(z)-\delta(z-\pi R)+\partial_z\sigma\}$. Note the bulk term in \eqref{eq:hirayo}, which comes from the non-trivial warp-factor. This term was neglected in the final result of \cite{hira}, which led to the wrong conclusion that SUSY is broken.


\section{Charged Hypermultiplets and Localisation}\label{sec:charged_hypers}

In this section we discuss the consequences of introducing hypermultiplets charged under the U(1)$_{FI}$. To be concrete let us consider in addition to the setup we had before a physical hypermultiplet $(H,~H^c)$ with charge $q_1=1$ (we absorb the charge in the gauge coupling $g_1$). Here, the chiral superfield $H$ will be taken to be even while $H^c$ is odd. One consequence of this is that the scalar component of the even compensator chiral superfield is now a function of $\mcA_H$ and $\mcA_H^c$, the scalar components of $H$ and $H^c$:
\be
                 h=e^{3\sigma/2}\kappa^{-1}\left\{ 1+\ka^2\left(|\mcA_H|^2+|\mcA_H^c|^2\right) \right\}^{\frac{1}{2}}+\theta^2 e^{5\sigma/2}F_h.
\ee
In addition there are new couplings involving $H$ and $H^c$:
\be\begin{split}
          \mcL_H=  2\int d^4\theta &\,{\mathbb W}_y\,\left( H^+ e^{-g_1 {V^1}}H + H^{c+} e^{g_1 {V^1}}H^c\right)\\
                 & -2\int d^2\theta H^c \left( \partial_y-g_1\Sigma^1 \right)H+\textup{h.c.}  
\end{split}\ee 
This leads to a new set of BPS conditions. From the conditions $F^c_h=0=F_H^c=F_H$ we get 
\be\label{eq:BPS_F_1}
               \left[\partial_y-\frac{e^5_y}{2}\epsilon (y) g_0 M^0\right]e^{3\sigma/2}\left\{ 1+\ka^2\left(|\mcA_H|^2+|\mcA_H^c|^2\right) \right\}^{\frac{1}{2}}=0,
\ee
\be\label{eq:BPS_F_2}
        \left[\partial_y-\frac{e^5_y}{2}g_1 M^1\right]e^{3\sigma/2}\mcA_H=0,    
\ee
and
\be\label{eq:BPS_F_3}
        \left[\partial_y+\frac{e^5_y}{2}g_1 M^1\right]e^{3\sigma/2}\mcA_H^c=0,    
\ee
while from $D^I=0$ we obtain (instead of \eqref{eq:BPS_D_without})
\be\label{eq:BPS_D}
                \partial_y e^{2\sigma}\mcN_J=2M_5^3 \,e^{2\sigma} e^5_y\, \epsilon(y)f_J (\mcA),
\ee
where 
\be
       f_J (\mcA)\equiv g_J\cdot \left\{\begin{array}{ll}  \left (1+\ka^2\left(|\mcA_H|^2+|\mcA_H^c|^2\right) \right) &, J=0,\\
                         \epsilon(y)\ka^2\left(|\mcA_H^c|^2-|\mcA_H|^2\right)& ,J=1.                         \end{array}\right.      
\ee

Now, we can combine eqs.\eqref{eq:BPS_F_1} to \eqref{eq:BPS_F_3} to get an equation for the warp-factor $\sigma(y)$, 
\be\label{eq:metric_hyper}
               \partial_y\sigma=e^5_y\frac{\epsilon (y)}{3M^3_5} \mcW, 
\ee
where the \emph{superpotential} $\mcW$ is defined as $\mcW\equiv M^3_5 f_I (\mcA) M^I$. Note that this very same equation follows by multiplication of eq.\eqref{eq:BPS_D} with $M^J$ upon the use of the constraint $\mcN=M_5^{\, 3}$, showing that one just needs to solve four of the above five equations. This constraint defines a 1-dimensional scalar manifold which can be parametrized by a single scalar $\phi$. In this way the scalars $M^I$ become functions of $\phi$. To obtain the equation of motion for $\phi$ we therefore have to contract eq.\eqref{eq:BPS_D} with $(\partial M^J/\partial\phi)$. After some manipulations, we get (using $\partial_y\mcN_J=\partial_{\phi}\mcN_J\,\partial_y\phi + \partial_{\epsilon}\mcN_J\,\partial_y\epsilon$)
\be\label{eq:BPS_scalar}
                g_{\phi\phi} \partial_y\phi = -e^5_y \epsilon(y) \frac{\partial\mcW}{\partial\phi} + \half\frac{\partial M^J}{\partial\phi}\frac{\partial\mcN_J}{\partial\,\epsilon}{\Big|}_{\phi}\partial_y\epsilon,
\ee
where we introduced the sigma-model \emph{metric}, $g_{\phi\phi}(\phi)$, defined by
\be
       g_{\phi\phi}(\phi)=-\half\mcN_{IJ}\frac{\partial M^I}{\partial\phi} \frac{\partial M^J}{\partial\phi}.        
\ee
Note that eq.\eqref{eq:BPS_scalar} is independent of the way we choose to parametrize the very special manifold. In particular we can take $\phi$ to be an \emph{even} scalar. This choice has the property that the second term at the r.h.s. of \eqref{eq:BPS_scalar} vanishes, and we get
\be\label{eq:BPS_scalar_even}
             g_{\phi\phi} \partial_y\phi = -e^5_y \epsilon(y) \frac{\partial\mcW}{\partial\phi}.     
\ee

\paragraph{Solutions of the BPS equations.} Let us now discuss the solutions of this new set of BPS equations. The first observation we make is that by integrating eq.\eqref{eq:BPS_D} over the whole extra dimension we obtain the constraint
\be\label{eq:constraint_1}
               \oint dy\,  e^{2\sigma}e^5_y (|\mcA_H|^2-|\mcA_H^c|^2)=0.
\ee
On the other hand, from eq.\eqref{eq:BPS_F_3} and the fact that $\mcA_H^c$ is odd, one gets that $\mcA_H^c=0$. Otherwise eq.\eqref{eq:BPS_F_3} would have singularities at the branes positions. It then readily follows that also $\mcA_H=0$, and we are back to the case discussed in section \ref{sec:BPS_FI_Terms} so that $M^0$ and $M^1$ are given by eqs.\eqref{eq:VevM^1} and \eqref{eq:VevM^0}, and the warp-factor is the one given in that section. 

Less trivial solutions, i.e. with non-vanishing hyperscalar VEVs, are possible if we add a second (bulk) hypermultiplet, $({\hat H},{\hat H}^c)$, with opposite charge, $q_1=-1$. While the odd hyperscalars are still vanishing, $\mcA^c={\hat\mcA}^c=0$, the constraint \eqref{eq:constraint_1} now gets replaced by
\be
              \oint dy\,  e^{2\sigma}e^5_y (|\mcA_H|^2-|{\hat\mcA}_H|^2)=0, 
\ee
which allows for non-trivial profiles for $\mcA_H$ and ${\hat\mcA}_H$. In this case, even if $g_0=0$, the metric will be warped, as follows from eq.\eqref{eq:BPS_F_1}:
\be\label{eq:warp_g0}
               e^{3\sigma}=\frac{c_0}{1+\ka^2(|\mcA_H|^2+|{\hat\mcA}_H|^2)}.
\ee
Note that we can obtain some additional knowledge about the solutions to the BPS equations by integrating eq.\eqref{eq:BPS_D} for $J=1$ over a small neighbourhood of the fix-point branes. In this way we learn that 
\be
                 M^1=\frac{\xi\epsilon(y)}{[1+(\ka\xi)^2]^{1/3}}+\psi,
\ee
where $\psi$ vanishes on the branes. This means that the value of $M^1$ near the branes is solely determined by the stength of the FI term. 

Let us solve the BPS equations for the case with $g_0=0$ and non-trivial profiles of the even hyperscalars. To parametrize the 1-dimensional very special scalar manifold we introduce an even scalar $\phi$ in the following way:
\be
               M^1(\phi)= \ka (\xi+\phi)\epsilon(y)M^0(\phi),
\ee
\be
               M^0(\phi)= \frac{\ka^{-1}}{[1+(\ka\xi)^2-(\ka\phi)^2]^{1/3}}.
\ee
We will have to resort to some approximation. We thus assume that $\ka|\phi|\ll 1$ and get:
\be\label{eq:phi_approx}
              \partial_y\phi\simeq e^5_y \epsilon(y)g_1(|\mcA_H|^2-|{\hat\mcA}_H|^2)[1+(\ka\xi)^2]^{2/3},
\ee
while from eq.\eqref{eq:BPS_F_2} (and a similar equation for ${\hat\mcA}_H$) we obtain, 
\be\label{eq:sol_hyp_scalr}
                  |\mcA_H|^2\simeq|a|^2\,\exp{(e^5_yg_1r(y))},\qquad  |{\hat\mcA}_H|^2\simeq|\hat a|^2\,\exp{(-e^5_yg_1r(y))},
\ee
where we chose a gauge with constant $e^5_y$, and introduced $r(y)\equiv \int^y_0 dy\, M^1$. In the bulk ($0<y<y_{\pi}$), from eq.\eqref{eq:phi_approx}, we obtain the following equation for $r(y)$:
\be\label{eq:mech}
                    \partial^2_y r=\frac{d}{dr}\left[|A|^2\cosh(e^5_yg_1(r-{\bar r})) \right],
\ee
where $|A|^2=2|a||\hat a|[1+(\ka\xi)^2]^{1/3}$, and ${\bar r}=(e^5_y g_1)^{-1}\ln|{\hat a}/a|$. 

Eq.\eqref{eq:mech} has a rather simple interpretation as being the equation of motion of a particle in a inverted cosh potential. The FI terms set boundary conditions at the two branes, $y=\{0,y_{\pi}\}$, which correspond, in the mechanical analogon, to fixing the start and end velocities: $\partial_y r(0)=\partial_y r(y_{\pi})=\xi[1+(\ka\xi)^2]^{-1/3}$. In addition, the initial position is $r(0)=0$, by definition. The fact that we have 3 boundary conditions implies that one of the parameters, $|A|$ or $\bar r$, is fixed by the other, the value of the FI terms and the size of the extra dimension . For special values of these parameters it is possible to solve eq.\eqref{eq:mech} analytically, and in this way to obtain the corresponding warp-factor. In particular, for $2|A|\cosh(g_1e^5_y{\bar r}/2)=|\xi|[1+(\ka\xi)^2]^{-1/3}$ we get
\be\label{eq:r_solution}
                    \exp\left(-\tfrac{1}{2}e^5_y g_1(r(y)-{\bar r})\right)=\frac{1+\tan\left(\frac{\xi}{4|\xi|}|A|g_1e^5_y(y_{\pi}-2y)\right)}{1-\tan\left(\frac{\xi}{4|\xi|}|A|g_1e^5_y (y_{\pi}-2y)\right)},
\ee
where we used 
\be\label{eq:boundary_con}
                \left|\frac{\hat a}{a}\right|= \tan^2 \left( \frac{\pi}{4}+\frac{\xi}{4|\xi|}|A|g_1e^5_yy_{\pi}\right),
\ee
which follows from the boundary condition at $y=y_{\pi}$. To obtain the warp-factor and hyperscalars profiles, we can use eqs.\eqref{eq:warp_g0} and \eqref{eq:sol_hyp_scalr}. We ilustrate our findings in Fig.1, with plots for a specific choice of the parameters. 

\begin{figure}
\resizebox{0.47\textwidth}{!}{
  \includegraphics{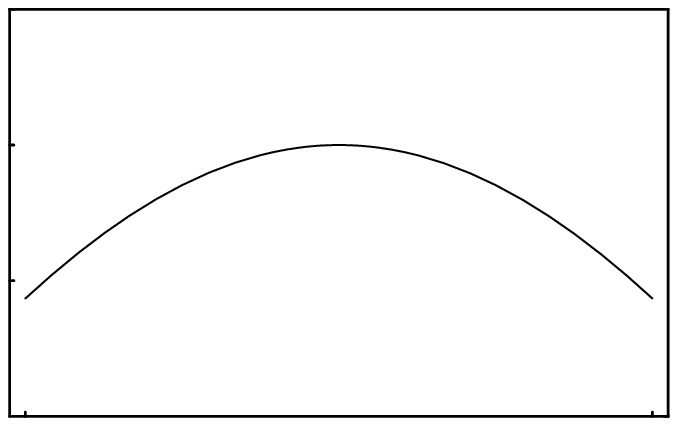}
}
\put(-210,2){$0$}
\put(-115,0){$y$}
\put(-35,3){$y_{\pi}$}
\put(-165,90){$e^{3\sigma}$}
\put(-222,88){$1$}
\put(-230,10){$0.8$}
\resizebox{0.47\textwidth}{!}{
  \includegraphics{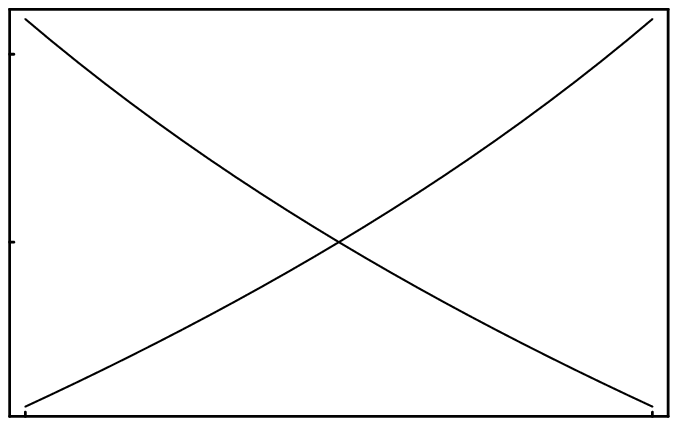}
}
\put(-210,2){$0$}
\put(-115,0){$y$}
\put(-35,3){$y_{\pi}$}
\put(-170,110){$\ka|{\hat\mcA}_H|$}
\put(-90,110){$\ka|{\mcA}_H|$}
\put(-230,115){$1.3$}
\put(-222,60){$1$}
\caption{Profiles of $\exp(3\sigma)$ and $|\mcA_H|$, $|{\hat\mcA}_H|$, for $g_1\xi>0$ and the parameters $|\hat a|=1.36 \ka^{-1}$, $|a|=0.74 \ka^{-1}$.}
\label{fig:1}       
\end{figure}

Perhaps the most salient feature of these solutions, and without the particular assumption we made above, is the fact that they correspond to vacua with the same \emph{negative tension} in both branes. This can be recognized from eq.\eqref{eq:metric_hyper} by noting that $\partial_y\sigma(0^+)=-\partial_y\sigma(y_{\pi}^-)>0$. To show this, we use again the mechanical analogon: since at the boundaries the \emph{velocities} are equal, the potential must also be the same. This implies that $r(y_{\pi})=2{\bar r}$. From eq.\eqref{eq:sol_hyp_scalr} it follows then that $|\mcA_H(0^+)|=|{\hat\mcA}_H(y_{\pi}^-)|$ and $|{\hat\mcA}_H(0^+)|=|\mcA_H(y_{\pi}^-)|$, and therefore we obtain
\be
               \partial_y\sigma(0^+)=-\partial_y\sigma(y_{\pi}^-)=\xi\ka^2(|\hat a|^2-|a|^2)\,\frac{e^5_y g_1}{3}[1+(\ka\xi)^2]^{-1/3}>0. 
\ee
The origin of these negative brane tensions is simple to understand. In each brane, the FI terms induce localised mass terms for both hyperscalars, which have the same magnitude but opposite sign. The positive mass repulses the corresponding hyperscalar from the brane while the hyperscalar with negative mass is attracted. This clearly has the net effect of producing negative tensions at both branes. Because of this, we expect the zero-mode of the graviton to be localised not on (one of) the branes but in the bulk. All these interesting features disappear for a vanishing FI term ($\xi=0$) since in this case the solutions are trivial: $|\mcA_H|=|{\hat\mcA}_H|=const.$, and $\phi=\sigma=0$. This is straightforward to show for the special solutions above, and can be proved in the general case using eqs.\eqref{eq:BPS_F_2}, \eqref{eq:BPS_F_3} and \eqref{eq:BPS_scalar_even} to show that in absence of FI terms $\phi$ is a monotone function of $y$. Since we must have $\phi(0)=\phi(\pi R)=0$ it follows that $\phi(y)=0$.


\paragraph{(De)localisation of hypermatter.} Let us now see the consequences of the stepwise VEV of $M^1$. In the case of rigid SUSY one knows that a hypermultiplet charged under the U(1)$_{FI}$ can get localised \cite{groot02, martiFI}. This is due to the fact that it gets an odd mass. Here the same happens. The Lagrangian includes a term
\be
                           -2\int d^2\theta { H}^c \left[ \partial_y-\half g_1 e^5_y M^1(y)\right]{ H}+\textup{h.c.}
\ee 
This shows that if there is a hyperscalar KK zero-mode $f_0(y)$, it must satisfy
\be\label{eq:zero_mode}
                 \left[ \partial_y-\half  e^5_y g_1M^1(y)\right]e^{3\sigma/2}f_0(y)=0.     
\ee

In the case that in the vacuum the physical hyperscalars vanish, $\mcA_H=0={\hat\mcA}_H$, the solution is rather simple to obtain:
\be
                f_0\propto \exp[\ka\xi g_1/g_0-1]3\sigma/2.
\ee
For $\xi g_1=0$ the localisation is due only to the warped geometry, while for $\xi g_1\neq 0$ the FI terms induce an additional amount of localisation in the same brane or localize the hyperscalar towards the other fix-point brane. Note that the best coordinate to evaluate this effect of (de)localisation is $y$, not $z$. In terms of $y$ the kinetic term of the zero-mode is already canonically normalized.

In the case with hyperscalars developing non-zero VEVs, the solution to eq.\eqref{eq:zero_mode} is just proportional to those VEVs. In the example we studied above, the scalar $\mcA_H$ is localised near one of the branes, ${\hat\mcA}_H$ near the other. The same happens with the zero modes.

\vs{0.5cm}
 
\hs{-0.7cm}{\bf Acknowledgments}
 
\vs{0.2cm}  
\hs{-0.7cm} The research of F.P.C. is supported by Funda\c c\~ ao para a Ci\^ encia e a Tecnologia (grant   SFRH/ BD/4973/2001).

\end{document}